\documentclass[jcp,aps,10pt,twocolumn,preprintnumbers]{revtex4-1}
\usepackage{graphicx}
\usepackage{amsmath, amsthm, amssymb} 
\begin{document} 

\title{Tuning effective interactions close to the critical point in colloidal suspensions}

\author{Nicoletta Gnan}
%\email{nicoletta.gnan@roma1.infn.it}
\affiliation{Dipartimento di Fisica and CNR-ISC, Universit\`{a} di Roma ``La Sapienza'', 
Piazzale A. Moro $2$, $00185$ Roma, Italy}

\author{Emanuela Zaccarelli}
\affiliation{Dipartimento di Fisica and CNR-ISC, Universit\`{a} di Roma ``La Sapienza'', 
Piazzale A. Moro $2$, $00185$ Roma, Italy}

\author{Francesco Sciortino}
\affiliation{Dipartimento di Fisica and CNR-ISC, Universit\`{a} di Roma ``La Sapienza'', 
Piazzale A. Moro $2$, $00185$ Roma, Italy}

\date{\today}

\begin{abstract}
We report a numerical investigation of two colloids immersed in a critical solvent, with the aim of
quantifying the effective colloid-colloid  interaction potential.  By turning on an attraction between the
colloid and the solvent particles we follow the evolution from the case in which the solvent density
close to the colloids changes from values smaller than the bulk to values larger than the bulk. We thus effectively implement the so-called $(+,+)$ and $(-,-)$ boundary conditions defined in field theoretical approaches focused on the description of  critical Casimir forces. We find that the effective potential
at large distances decays exponentially, with a characteristic decay length compatible with the bulk
critical correlation length, in full agreement with theoretical predictions.  We also investigate the case
of $(+,-)$ boundary condition, where the effective potential  becomes repulsive.  
Our study  provides a guidance for a design of the interaction potential which can be exploited to control the stability of colloidal systems. 
\end{abstract}

\maketitle

\section{Introduction}
Interactions between colloidal particles in dispersions depend, beside the colloid coordinates, on the degrees of freedom of the solvent and of the co-solutes.  When a clear separation in time and space scales between solvent molecules and
colloidal particles exists, it is possible to integrate out the solvent (and co-solute)
degrees of freedom and derive an effective  potential that describes
the interaction between  colloids~\cite{Likos}. A well known case is the depletion interaction derived long time ago by Asakura and Oosawa~\cite{AO} which have shed light on the role of entropic forces on the phase diagrams of colloidal suspensions.  Depletion interactions arise whenever small additives,  such as polymers or surfactants,  are added to colloidal dispersions; when two colloids are closer than the typical size of the co-solute, 
the latter is excluded by a  ''depletion'' region in between the two colloids. As a result, a pressure gradient originates, giving rise to a short-range entropy-driven attractive effective potential -- the depletion potential.
Today we know that depletion forces can be strong enough to induce colloidal phase separation~\cite{LekNat416,LekEPL20,Dijkstra}, or 
to enhance the stability of the crystalline phase, as in the case of proteins~\cite{FrenkelSCI26,PiazzaCOCIS}.

An interesting effective potential is the one arising when  two colloids are dispersed in a critical fluid.
Close to the critical point, thermal fluctuations of the order parameter 
are correlated over length-scales that are much larger than the solvent size and their properties become
 independent of the microscopic behavior of the  system, i.e. they are \emph{universal}. 
 The confinement of the order-parameter fluctuations in between the  two colloids  give rise to long-range effective forces, named  critical Casimir forces~\cite{Kerch}. 

In the last decade direct experimental evidence of such forces has been provided~\cite{hertleinnat451}. 
Theoretical investigation of critical Casimir forces, which started with the pioneering work of 
Fisher and de Gennes~\cite{Fisher} back in 1978,  has provided accurate predictions for the
radial dependence of the effective potential and the associated scaling properties. 
Such studies, capitalizing on the universality aspect of the problem, have been mostly focusing on
Ising model calculations~\cite{VasilyevEPL,MaciolekPRE79,MohryPRE2010,VasilyevPRE88} and, more recently,  on a field-theory approach for a classical binary mixtures~\cite{gambassipre80,MohryArXiv1,MohryArXiv2,Buzzaccaro,PiazzaJPCM23}. The main result of these studies is the explicit derivation of the critical Casimir potential for a  solvent confined between the surfaces of two large colloids:

\begin{equation}\label{eq:Phi_z}
\beta \Phi(z)=\frac{\sigma_c}{z}\Theta(z/\xi).
\end{equation} 

where $\sigma_c$ is the colloid diameter and $z$ is the distance between the surfaces of the two colloids.  The scaling function $\Theta(z/\xi)$ depends on the distance from the critical point (via the correlation length $\xi$), on the bulk universality class of the solvent and on the boundary conditions (BC) imposed by the colloidal surface properties.
It has been demonstrated~\cite{gambassipre80} that  for $z>>\xi >> \sigma_s$ (the latter being a measure of the solvent size), the behavior of the scaling function in Eq.~(\ref{eq:Phi_z}) is

\begin{equation}\label{eq:Theta_minus}
\Theta(z/\xi)_{(\pm,\pm)} (z/\xi\gg 1) =  \pi A_{(\pm,\pm)} (z/\xi) e^{-(z/\xi)}
\end{equation}

\noindent where the $(\pm,\pm)$ signs are related to  different BCs, i.e., to the different  absorption preferences of the confining surfaces with respect to the solvent: $(+,+)$ or $(-,-)$ corresponds to symmetric BC and  $(+,-)$ (or equivalently $(-,+)$) to asymmetric BC. Experimental results have shown that it is possible to generate repulsive and attractive critical Casimir forces between a colloid and a substrate by modifying the BC of the substrate~\cite{hertleinnat451}. Moreover it has been shown that is possible to continuously  tune the force from attractive to repulsive  by producing a gradient
in the physico-chemical properties of the substrate~\cite{NellenEPL88}. This can be exploited for inducing orientationl-dependent effective forces in colloids close to chemically patterned surfaces~\cite{GambassiSM7,soykaprl101}. Critical Casimir forces have also been observed when the
critical behavior of the host medium refers not to the solvent itself but  to the presence
of small interacting co-solutes added in solution.  If the inter-cosolutes interaction becomes
strong enough, a phase separation (the analog of the gas-liquid) takes place in which the co-solutes partition themselves  in two phases of different concentration.  Close to the corresponding critical point, the critical fluctuations in co-solute concentration generate critical Casimir forces. Such mechanism has been exploited in the experimental study of
Buzzaccaro and coworkers~\cite{Buzzaccaro}, in which PMMA colloidal particles are dispersed in an interacting micellar solution.
In the same work, the authors presented an interesting connection between depletion forces generated by the presence of the micelles far from the critical point and Casimir forces generated by the critical fluctuations close to the micellar critical point.

Most of the theoretical and numerical investigations  of Critical Casimir forces have been based on studies
of lattice models, exploiting the universality properties of the phenomenon. In a recent study~\cite{GnanSM8}, 
we have reported  a numerical evaluation of the effective interaction potential between two
spherical hard-sphere colloidal particles immersed in a critical depletant, modeled as short-ranged square-well attractive potential, with the aim of exploring how the interaction potential changes from the 
depletion shape occurring far from the critical point, to the universal shape induced by critical fluctuations close to the depletant critical point.  We have confirmed the critical nature of the effective potential close to the critical point
by showing that  the quantity $\xi$ entering in the effective potentials  (see Eq.~\ref{eq:Theta_minus}) is consistent with the bulk correlation length of the critical co-solute extracted from the static structure factors.  Interestingly enough, we have found that the strength of the effective potential
between the two hard-sphere colloids (when the colloid-depletant interaction is only controlled by excluded volume, and hence in the $(-,-)$ BC class) is sufficient to drive bulk phase separation of the colloidal solution well before the
critical region is approached.   

In this article we explore  the changes in the effective potential when the colloid-cosolute interaction is continuously modified from hard-core repulsion to strong attraction, continuously moving from the $(-,-)$ to the $(+,+)$ BC.
We also explore the interesting case in which $(+,-)$ BC are present, i.e. the case of two colloids interacting
in different ways with the cosolutes, again interpolating between the $(-,-)$ and the $(+,-)$ limits on changing  (this time only for one of the two colloids) the
colloid-cosolute interaction strength.   We calculate numerically the effective potential both at a high $T$, where critical phenomena are absent as well as close to the co-solute critical point.

\section{models and methods}
The total colloid-colloid interaction potential $\phi_{CC}$  results from the sum of the bare colloid-colloid interaction $V_{CC}$ and of the effective potential $V_{eff}$ arising from the integration of the solvent and cosolute degrees of freedom,
\begin{equation}
\phi_{CC}(r_{12})=V_{CC}(r_{12})+V_{eff}(r_{12})
\end{equation}

We model  $V_{CC}$    as a excluded volume interaction  between two colloids of size $\sigma_c$
\begin{equation}
V_{CC}(r_{12})=\begin{cases} \infty, & r_{12}<\sigma_c \\ 
 0 & r_{12}\geq \sigma_c
\end{cases}
\end{equation}

Inspired by the work of Buzzaccaro and coworkers~\cite{Buzzaccaro}, we model the critical medium as a
fluid of interacting co-solutes dispersed in an implicit solvent.  The co-solutes, 
of size $\sigma_s$  interact via a pairwise square-well potential (SW) 
\begin{equation}
V_{SS}(r_{ij})=\begin{cases} \infty, & r_{ij}<\sigma_s \\ 
-\varepsilon_s, & \sigma_{s}\leq r_{ij}< (1+\delta) \sigma_s \\ 
 0 & r_{ij}\geq (1+\delta)\sigma_s\\
\end{cases}
\end{equation}

\noindent  where $\varepsilon_s$ controls the strength of the interaction and $\delta $ the relative 
(respect to $\sigma_s$) width of the well. $\sigma_s$ and $\varepsilon_s$ are chosen as unit of length and energy. The temperature T is measured in units of $\varepsilon_s$. The co-solute is characterized by a gas-liquid critical point located at $(T_c=0.478,\phi_c=0.25)$~\cite{largojcp128}, where $\phi_c=(\pi/6)\rho_c \sigma_s^3$ is the critical packing fraction of the co-solute and $\rho_c$ is its number density.  For this model, it has been shown~\cite{GnanSM8} that
the correlation length and the susceptibility extracted from the static structure factors close to the critical point
diverge with a power-law with the respective Ising critical exponents.   The size ratio $q\equiv \sigma_s/ \sigma_c$ between the co-solute and the colloids is fixed at $q=0.1$.

To evaluate the effective potential, we perform Monte Carlo simulations of two colloids in a fluid of 
co-solute particles, in the canonical ensemble at fixed $T$ and $\rho_s$ in a parallelepiped shape. 
The two colloids are constrained to move only along the $x-$axis, sampling only a
limited range of distances. Several overlapping relative distances windows are simulated, evaluating for each window $P(r)$, the probability of observing the two colloids at relative distance $r$.  Splicing together
the $P(r)$ evaluated in different windows provides an effective (and parallel) way for evaluating the entire 
$P(r)$. The logarithm of $P(r)$ is by definition the effective potential (apart from an overall constant which is fixed imposing $V_{eff}(\infty)=0$).  To minimize finite-size effects at the temperatures investigated, the dimensions of the box ($L_x$;$L_y$;$L_z$) are chosen in such a way that the surface-to-surface colloidal distance evaluated via the boundary conditions is more than twice the distance over which the effective potential goes to zero.
Moreover, along all directions, the solvent density profile reaches a constant  value on approaching the box boundaries.  Close to the critical point and along the critical isochore, the size of the box is $[L_x = 52;L_y = 26;L_z = 26]$, requiring 16000 co-solute particles. The bulk density is estimated a posteriori by calculating the local density far from the two colloids.

We also evaluate the co-solute density profile for different BC by selecting  a volume centered  along the $x-$axis  of transversal section  equal to $\sigma_s^2$ and we average the local density with a mesh of the order of $0.1 \sigma_s$. 

In this paper, when discussing the critical behavior, we will show results for effective potentials and density profiles evaluated at the critical packing fraction and at the reduced temperature $T/T_c=1.0251$, corresponding to a critical correlation length $\xi=2.5\sigma_s$~\cite{GnanSM8}.

\section{Results}

\subsection{$(-,-)$ BC}
In our previous work~\cite{GnanSM8} we have discussed the evolution of the effective potential when 
the interaction between the colloids and the solvent  $V_{CS}$ are treated as hard-spheres. In such condition, close to the colloid surface, the density of the solution is smaller than the average, effectively generating a $(-,-)$ BC.
We have shown that close to the critical point, $V_{eff}$  is long-ranged, signaling the onset of criticality. 
Its radial dependence is well described by the exponential decay of Eq.~\ref{eq:Phi_z}, with the same characteristic length $\xi$ of the bulk critical fluctuations. 

\begin{figure}[ht]
\centerline{\includegraphics[width=.9\linewidth]{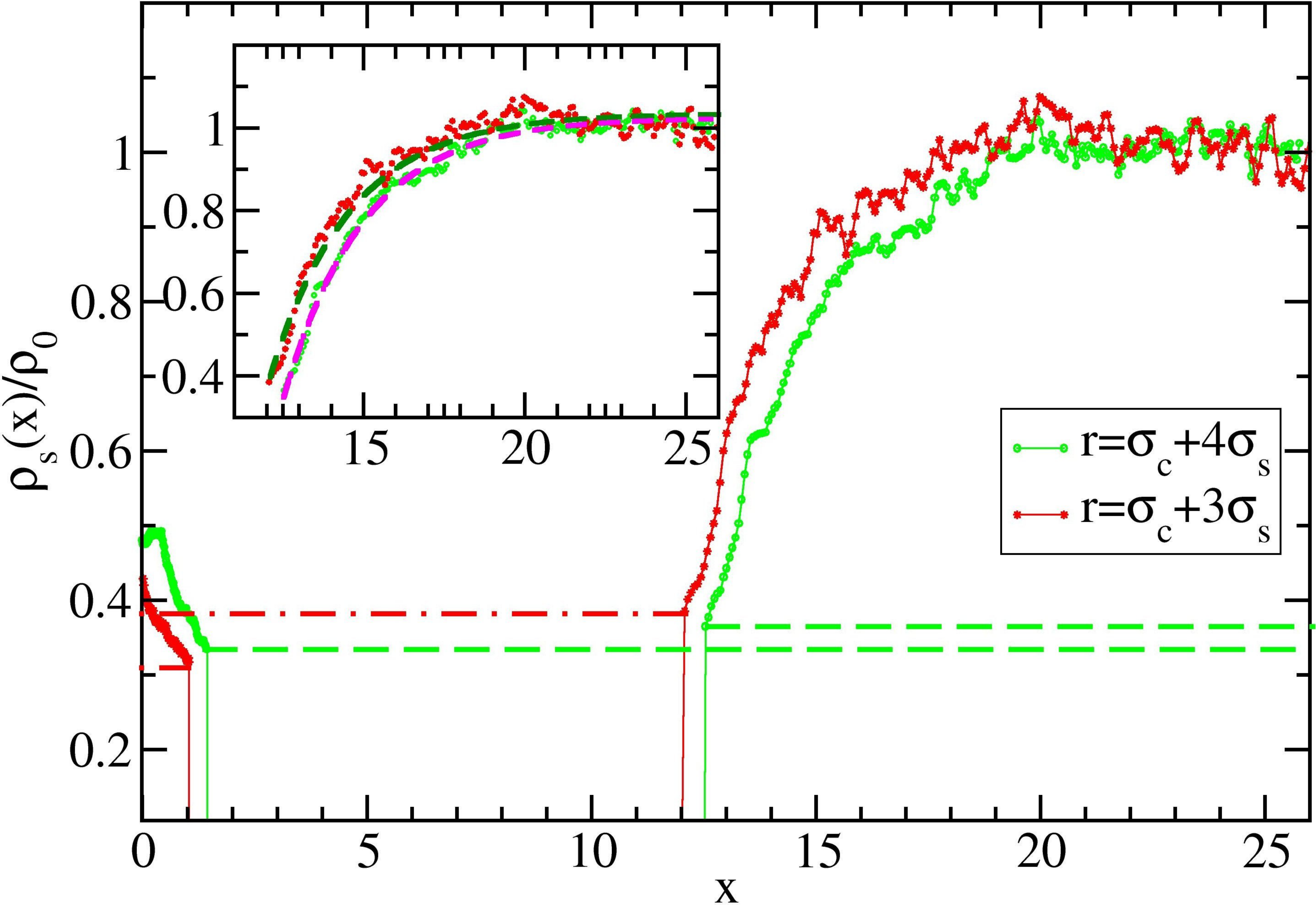}}
\caption{\label{fig1}
Normalized co-solute density profiles with respect to the bulk density $\rho_0$ along the horizontal $x$-axis at $T/T_c = 1.0251$ and at the critical density $\phi_c=0.2501$.  Co-solute particles interact with a short-range SW attraction while colloid-cosolute interaction is modeled with a HS repulsion.  The two profiles are evaluated for two different colloid-colloid distances, namely for $r=\sigma_c+3\sigma_s$ and $r=\sigma_c+4\sigma_s$, with $\sigma_c=10\sigma_s$. The box is centered at $x=0$ and the two colloids are centered at $\pm r/2$  along the $x$-axis. Being the profiles symmetric with respect to $x=0$ only positive $x$ are shown here. Note that each colloid excludes the solvent in a region equal to $\sigma_c+\sigma_s$. The dashed lines highlight the mismatch in the density at contact between the inner and outer side of the colloid. The inset shows that the outer part of the density profile is well fitted by an exponential $\rho=\rho_0+A \exp(-r/\xi)$, where $\xi$ has been fixed to $\xi=2.5\sigma_s$, i.e. to the bulk critical correlation length value for $T/T_c=1.0251$ at $\phi_c$. In this case the fit parameter $A<0$.}
\end{figure}

To get more insight into the mechanism which drives attraction between colloids close to the critical point in the $(-,-)$ BC case, we show in Fig.~\ref{fig1} the co-solute density profile along the horizontal $x$-axis for two different relative colloid-colloid distances. Notice that the two colloids are always located symmetrically with respect to the origin. Thus the density profile is symmetric with respect to $x=0$. For this reason in Fig.~\ref{fig1} we show only the positive  
$x$ region. In between the two colloids, the density is significantly lower than the bulk density, a typical depletion effect.  Outside, the density relaxes toward the bulk value with an exponential decay, again controlled by the bulk critical correlation length.  
It is interesting to discuss the physical origin of the net attractive force between the two colloids. Since the co-solute colloid interaction is modeled via an hard-sphere potential, it is possible to prove that the effective force results
from the mismatch in the contact density along the two sides of each colloid. Fig.~\ref{fig1} shows that indeed, at contact, the density outside is slightly larger than inside. The figure also shows that the  mismatch decreases on increasing the relative distance between the two colloids in parallel with the decrease of the effective force.

\subsection{From the $(-,-)$ to the $(+,+)$ BC}

To drive the transition from $(-,-)$ toward $(+,+)$ BC we tune the strength of a 
short-range attraction in the  colloids-cosolute interaction $V_{CS}$.  
The attraction gives rise to an enhanced accumulation close to the surface of the colloids, at first compensating and then inverting the depletion characteristic effect of the excluded volume interaction~\cite{KaranikasJCP128,foffijpcb114}. 

 We  model the attraction via a pairwise SW potential

\begin{equation}
V_{CS}(r_{ij})=\begin{cases} \infty, & r<\sigma_{cs} \\ 
-\varepsilon, & \sigma_{cs}\leq r<  \sigma_{cs} + \Delta \sigma_s\\ 
 0 & r\geq \sigma_{cs} + \Delta \sigma_s\\
\end{cases}
\end{equation}

where $\sigma_{cs}=(\sigma_s+\sigma_c)/2$ and $\Delta=0.35$.
The width $\Delta \sigma_{s}$ has been chosen as a compromise between limiting the colloid-cosolute interaction to the nearest-neighbor shell and maximizing the volume over which 
cosolute bind to the colloid.    The parameter $\varepsilon$ is used as control parameter to drive the cross-over from the hard-sphere like behavior ($\varepsilon \rightarrow 0$) to the wetting case $\varepsilon/kT  \gtrsim 1$).  

\begin{figure}[ht]
\centerline{\includegraphics[width=.9\linewidth]{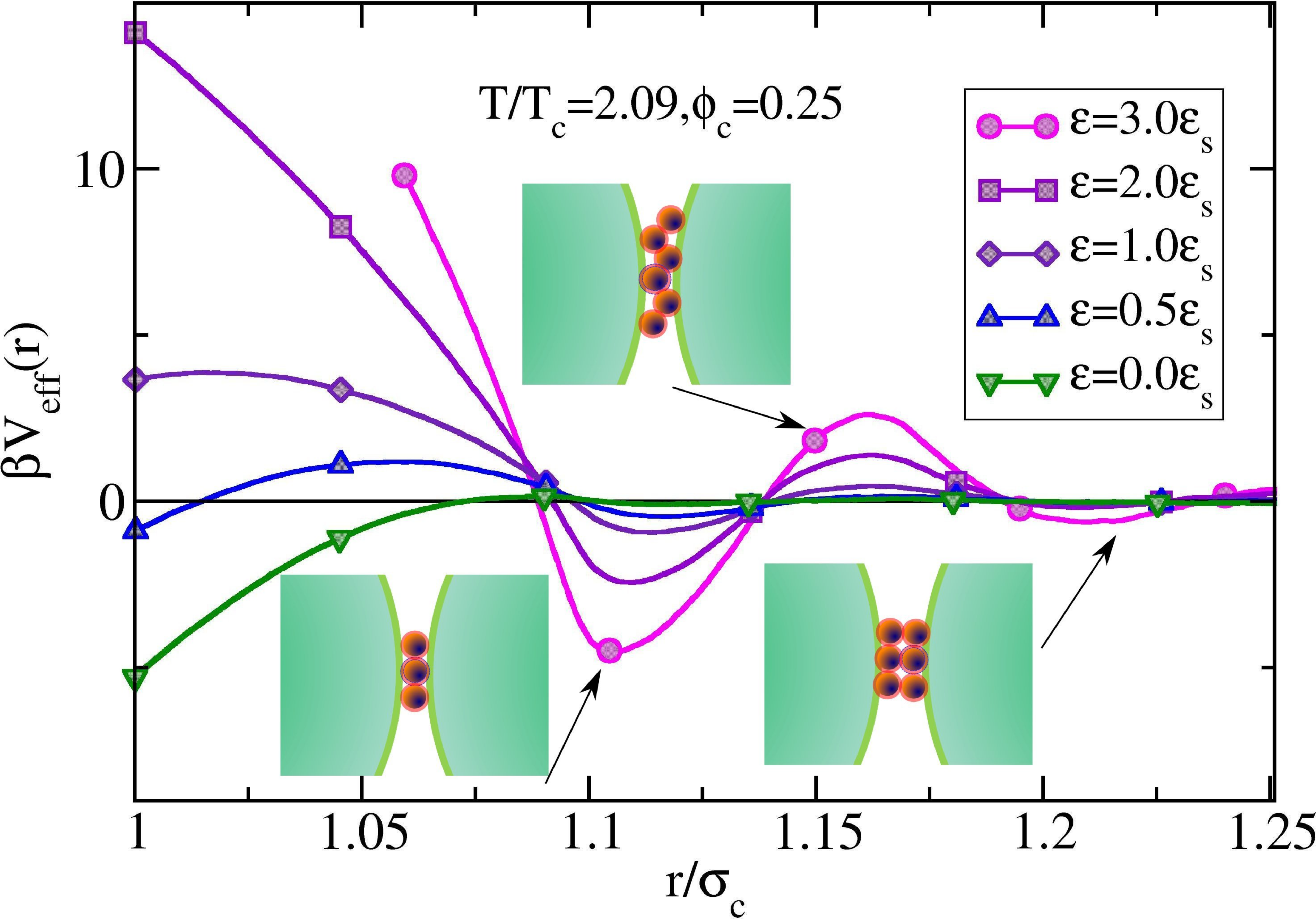}}
\caption{\label{fig2}
Effective potential for two large colloids interacting through a hard-core repulsion at $T/T_c=2.09$ and $\phi=\phi_c=0.25$. Colloids are in solution with co-solute particles interacting (among themselves) via a SW attraction of width $\delta\sigma_s=0.1\sigma_s$ and depth $\varepsilon_s$.
The two colloids interact  with co-solute particles via a SW attraction of width $\Delta\sigma_s=0.35\sigma_s$ and depth $\varepsilon$. 
Sketches in the figures represent particular configurations adopted by the co-solute  for different colloid-colloid distances. Such configurations give rise to maxima and minima in the potential. }
\end{figure}

We start by discussing the behavior of the effective potential at high $T$, where critical phenomena are not present. The evolution of the effective potential  upon changing $\varepsilon$ is shown in Fig.~\ref{fig2}. When $\varepsilon=0$,  the well known depletion
interaction potential is observed.  On increasing $\varepsilon$, the depletion attractive interaction is progressively weakened, and the potential at contact becomes repulsive. 
For  even large $\varepsilon$ values, the colloid becomes surrounded by a persistent layer
of co-solutes which extent the effective radius of the colloid, making it impossible
to attract a neighboring colloid for distances closer than $\sigma_c+\sigma_s$~\cite{GrestJCP104}.
Under such strong coupling conditions, the effective  potential acquires an oscillatory character, with minima originating from the preferential distances  allowing for an integer number of co-solute layers between the colloids.  The configuration associated to the
first minimum, called bridging, is the most energetically favorable since one co-solute particle is bonded with both  colloids. The other minima are related to particular configurations in which bridging is obtained by particle chains. A sketch of such situation is shown in Fig.~\ref{fig2}. The intermediate maxima occur when the co-solute particles (or particle chains) are not bonded to both colloids. For example the first maximum corresponds to a situation, also illustrated in Fig.~\ref{fig2}, in which a single co-solute particle cannot be bonded to both colloids, since $r>\sigma_s$.
 The evolution of the effective potential reported in Fig.~\ref{fig2}  clearly show how the minimum at contact progressively turns into a maximum,  and simultaneously a new minimum develops at the bridging distance. Intermediate values of $\varepsilon$ thus provide  a viable mechanism for contrasting depletion interaction  and favoring colloidal stability.  The case $\varepsilon=0.5$ is emblematic, since the 
minimum at contact has essentially disappeared while the new minimum at distance $1.1\sigma_c$  has not yet developed, such that the effective potential
is  never (in absolute value) significantly larger than $k_B T$.   For very large $\varepsilon$ values, $V_{eff}$  becomes again sufficiently intense to drive a colloid phase separation.

\begin{figure}[ht]
\centerline{\includegraphics[width=.9\linewidth]{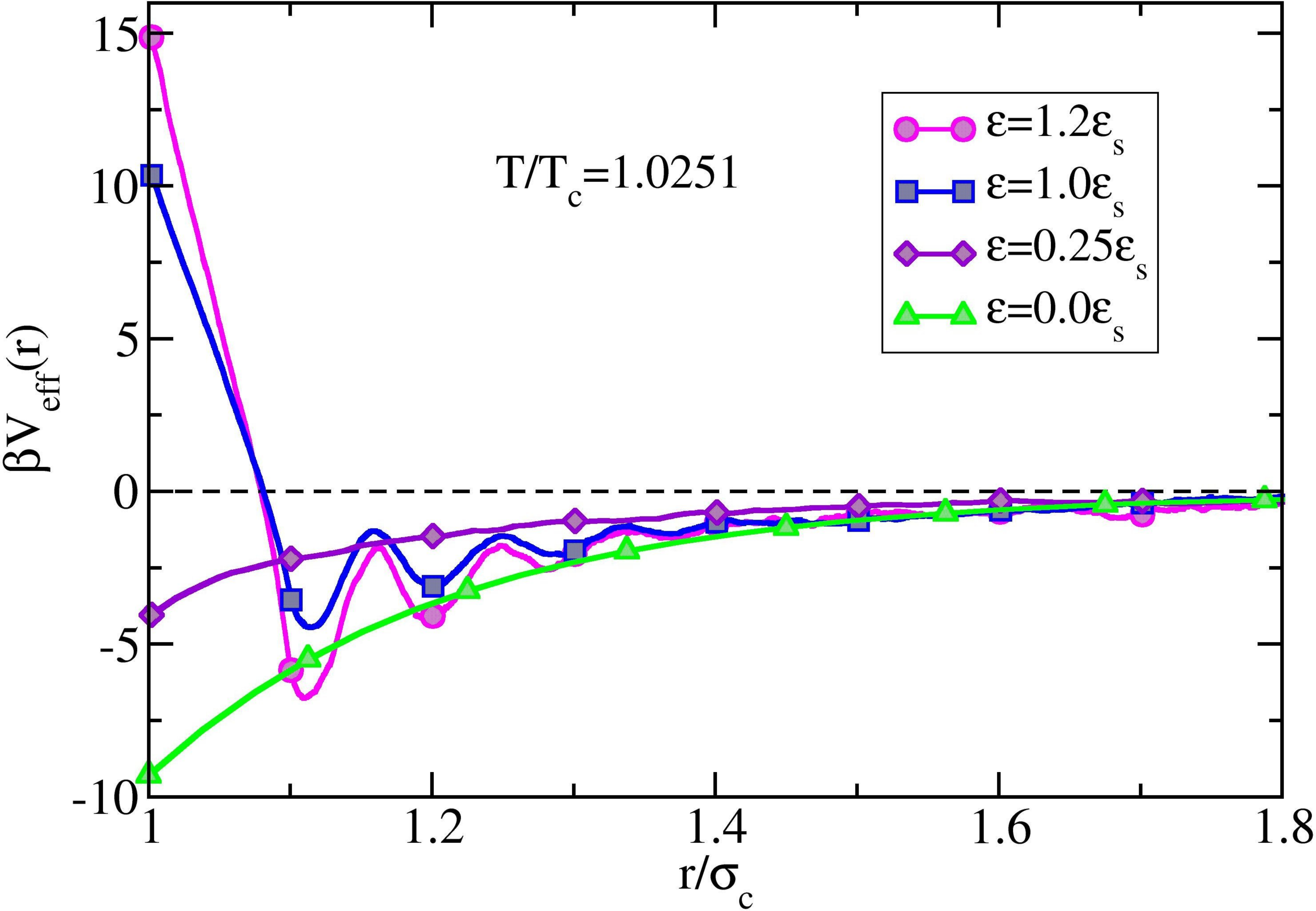}}
\caption{\label{fig3}
Effective potential of two large colloids close to the critical point of the co-solute ($T/T_c=1.0251$ and $\phi=\phi_c=0.25$). The two colloids interact  with co-solute particles via a SW attraction of width $\Delta\sigma_s=0.35\sigma_s$ and depth $\varepsilon$. The two colloids have therefore (+,+) BC. The increase of the attraction strength give rise to large oscillations associated to bridging effects as discussed in Fig.~\ref{fig2}.}
\end{figure}

\begin{figure}[ht]
\centerline{\includegraphics[width=.9\linewidth]{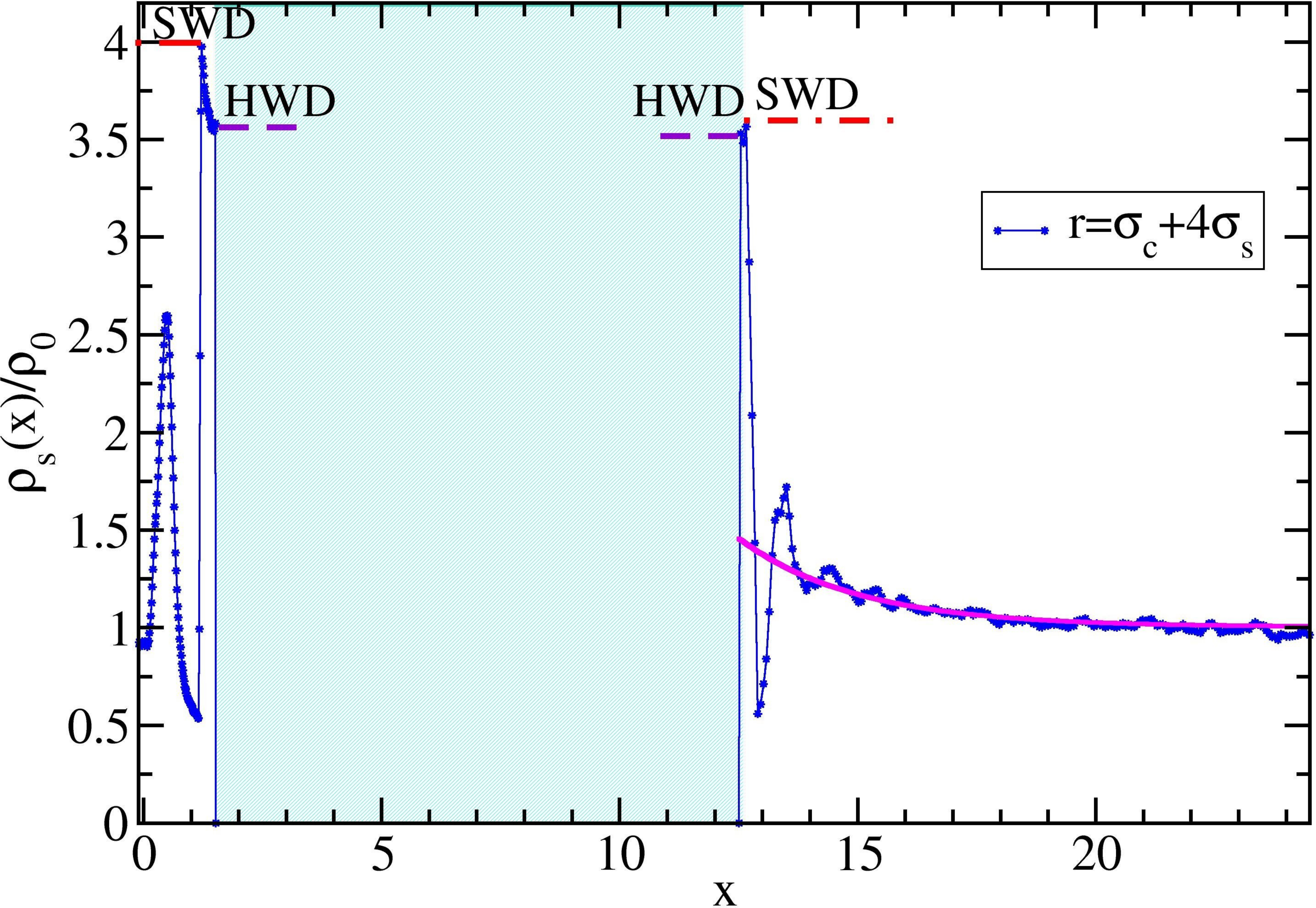}}
\caption{\label{fig4}
Normalized co-solute density profile  with respect to the bulk density $\rho_0$ along the horizontal $x-$axis at $T/T_c=1.0251$ and at the critical density $\phi_c=0.2501$.  Co-solute particles interact with a short-range SW attraction of width $\delta\sigma_s=0.1\sigma_s$ and depth $\varepsilon_s$ while colloid-cosolute interaction is modeled with a SW interaction of width $\Delta\sigma_s=0.35\sigma_s$ and depth $\varepsilon=1\varepsilon_s$. 
The box is centered at $x=0$ and the two colloids are centered at $\pm r/2$  along the $x$-axis.
Being the profile symmetric with respect to $x=0$ only positive $x$ are shown here. Note that each colloid excludes the solvent in a region equal to $\sigma_c+\sigma_s$. The dashed segments highlight the mismatch in the density at contact (HWD) and at the well boundary (SWD) between the inner and outer side of the colloid.The profile is associated to the effective potential in Fig. 3 (squares). As for Fig. 1, the criticality of the co-solute can be caught from the exponential decay of the density outside the colloids towards $\rho_s(x)/\rho_0=1$. The exponential fit shown in the figure (solid line), has been performed by fixing the exponential decay with the value of the bulk critical correlation length for the temperature and density investigated, i.e. $\xi=2.5\sigma_s$.}
\end{figure}

Close to the critical point, critical Casimir forces add to the standard effects described 
before.  Differently from the pure $(-,-)$ case, the presence of an
attractive interaction between the colloid and the solvent brings in two new features that are distinguishable for medium and large values of $\varepsilon$:
a repulsive contribution to the effective potential at short distances and
the presence of oscillations induced by the granularity of the cosolutes. 
The theory on critical Casimir forces predicts that at distances larger than $\xi$ the  effective potential develops an attractive exponential tail, so that the effective potential behaves the same for the  $(-,-)$ BC  and the  $(+,+)$ BC cases.  Fig.~\ref{fig3} shows how the effective potential evolves on increasing the value of  $\varepsilon$.  As expected the contact value becomes repulsive  but the long tail behavior remains always attractive and has the same exponential character of the  $\varepsilon=0$ case~\cite{GnanSM8}.  Even for large values  of $\varepsilon$,  where the same oscillations characterizing the high T effective potentials modulate the  shape,   the overall behavior can be represented by an exponential function. 
The density profile for the $(+,+)$ case is shown in Fig.~\ref{fig4}. Close to the colloidal surfaces a significant layering of the cosolutes is observed,  consistent with the presence of oscillations in $V_{eff}(r)$.  On the external sides, beyond the oscillations, the density decays again with an exponential shape controlled once more by $\xi$. 
In comparison with the $(-,-)$ case, the resulting interaction between the two particles is more difficult to visualize, since it arises from the competition between the density at contact  (in which the solvent pushes the colloid)  and the density the well boundary (where the solvent attracts the colloid). Indeed, in the case of a colloid of diameter $sigma_C$ interacting via square-well attraction with solvent particles, the pressure originates from the two points of discontinuity of the potential,
the colloid-cosolute hard-wall distance (HWD) and the square-well distance (SWD) $\sigma_cs+\Delta\sigma_s$ according to the expression~\cite{HansenMcDonald}:

\begin{eqnarray}\label{eq:pressure}
\frac{\beta P^{ex}}{\rho}&=&\frac{2}{3}\pi\rho [\sigma_{cs}^3 g(\sigma_{cs}^{+}) \nonumber \\
&-&(\sigma_{cs}+\Delta\sigma_s)^3 g(\sigma_{cs}+\Delta\sigma_s^{-})(1-e^{\beta u_0})].
\end{eqnarray}

\noindent where $g(\sigma_{cs})$ is the pair correlation function evaluated at the HWD and $g(\sigma_{cs}+\Delta\sigma_s)$ at the SWD. 
From Eq.~\ref{eq:pressure} one can notice that the density at the hard-wall generates a positive contribution to the pressure, while the density at the well boundary provides a negative contribution. Figure~\ref{fig4} shows that indeed the density at the SWD inner side (left side of Fig.~\ref{fig4}) is significantly higher than the density at the SWD outer side (right side of Fig.~\ref{fig4}), and it is responsible for the
resulting net attractive force between the two colloids. Indeed, the excluded volume contribution in this case would tend to
separate the two colloids, being the contact density outside slightly larger than the contact density inside.

\subsection{From the $(-,-)$ to the $(+,-)$ BC}

We now  discuss the situation in which one of the two colloids $(C_1)$ interact with the co-solute through a square-well attraction ($V_{C_1S}(r_{1i})$) while the other colloid $(C_2)$ experience only excluded volume interactions ($V_{C_2S}(r_{2i})$). More precisely

\begin{equation}
V_{C_1S}(r_{1i})=\begin{cases} \infty, & r_{1i}<\sigma_{cs} \\ 
-\varepsilon, & \sigma_{cs}\leq r_{1i}< \sigma_{cs}+\Delta \sigma_{s} \\ 
 0 & r_{1i}\geq \sigma_{cs}+\Delta \sigma_{s} \\
\end{cases}
\end{equation}

 and 

\begin{equation}
V_{C_2S}(r_{2j})=\begin{cases} \infty, & r_{2j}<\sigma_{cs} \\ 
 0 & r_{2j}\geq \sigma_{cs}.\\
\end{cases}
\end{equation}

\begin{figure}[ht]
\centerline{\includegraphics[width=.9\linewidth]{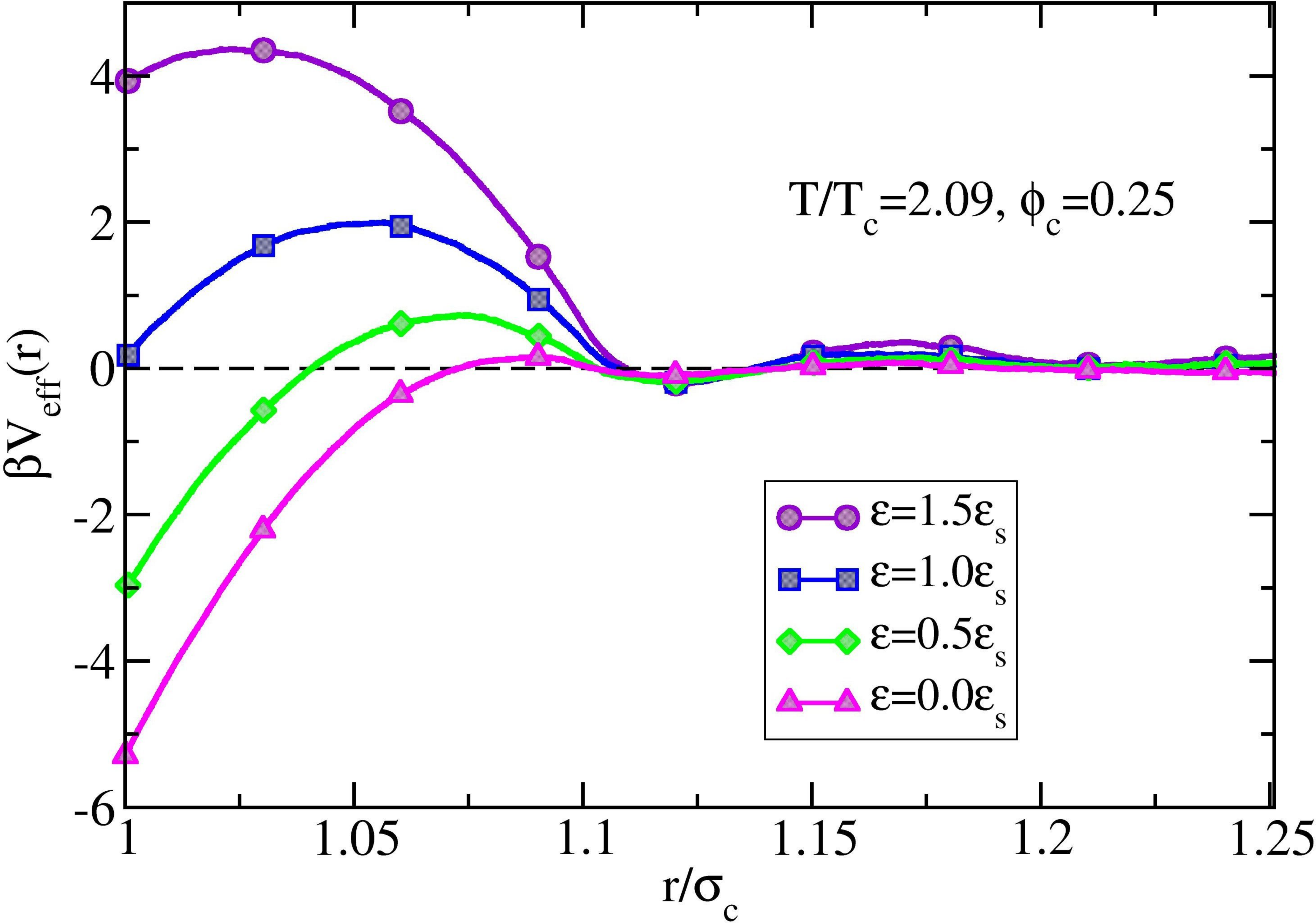}}
\caption{\label{fig5}
Effective potentials for two large colloids interacting through a hard-core repulsion at $T/T_c=2.09$ and $\phi=\phi_c=0.25$. Co-solute particles, in solution with the two colloids, interact through a SW attraction of width $\delta=0.1\sigma_s$ and depth $\varepsilon_s$.
One of the two colloids ($C_1$) interact with co-solute particles via a SW attraction of width $\Delta=0.35\sigma_s$ and depth $\varepsilon$. The other colloid($C_2$) has instead a HS repulsion with the co-solute.}
\end{figure}

Also in this case we start investigating a $T$ significantly larger than the critical one  for different interaction strengths $\varepsilon$. The results are shown in Fig.~\ref{fig5}.

The trend is similar to the one observed going from $(-,-)$ to the $(+,+)$ case; (i) the contact value of the effective potential
grows continuously on increasing  $\varepsilon$, progressively offsetting the original $\varepsilon=0$ depletion interaction.  Still, the strength of the repulsion is significantly smaller than the one observed in the  $(+,+)$ case.
(ii) only weak oscillations (with amplitude smaller than $k_BT$) characterize the radial dependence of the effective potential, signaling the absence of a strong layering of the cosolute between the colloids.  We notice that similar features have also been observed in the case of non-additive HS mixtures~\cite{LouisJPCM2001,RothPRE2001}.
 
\begin{figure}[ht]
\centerline{\includegraphics[width=.9\linewidth]{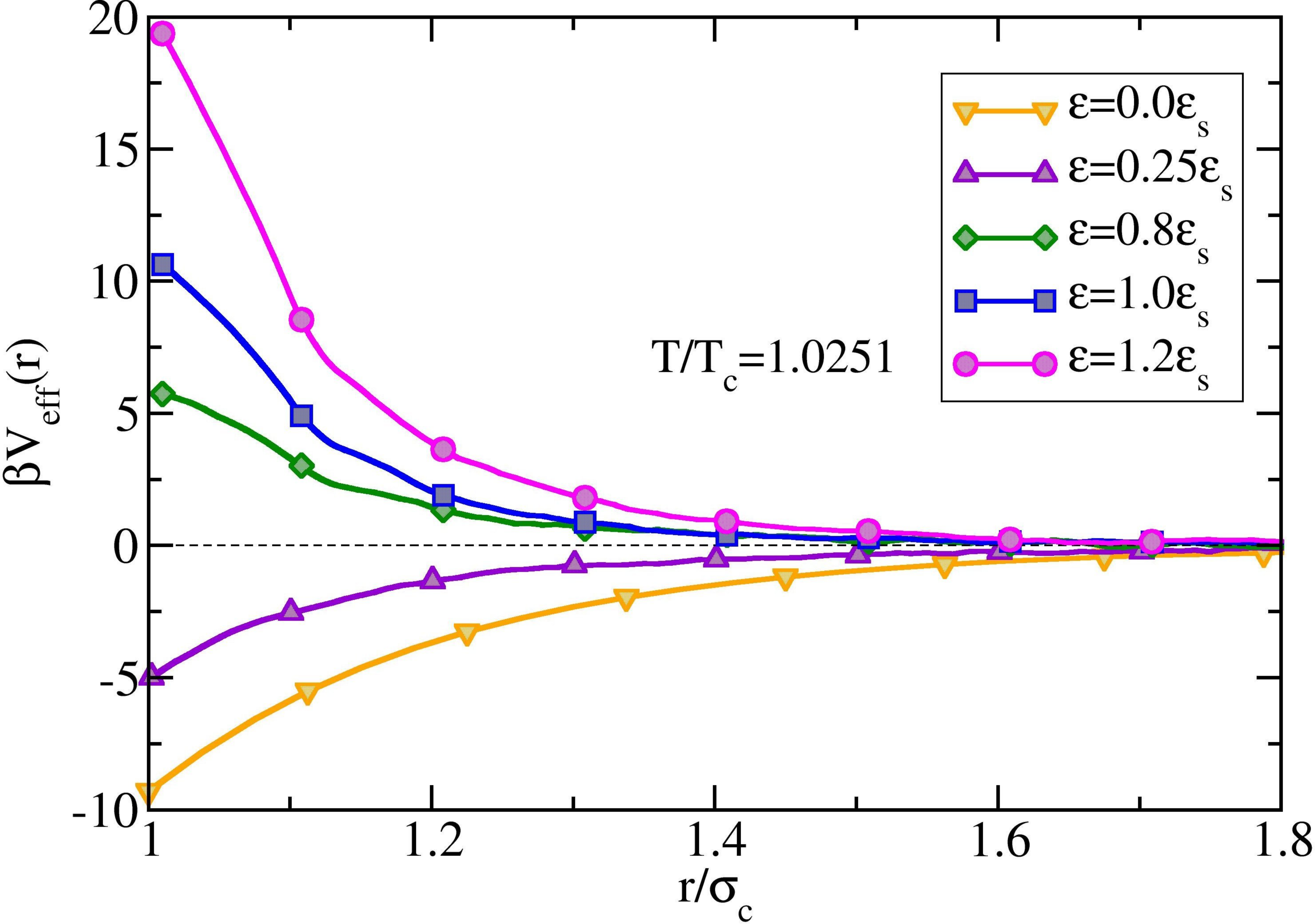}}
\caption{\label{fig6}
Evolution of the effective potential of two colloids close to the critical point of the co-solute ($T/T_c=1.0251$ and $\phi=\phi_c=0.25$) on increasing $\varepsilon$. One of the two colloids ($C_1$) interact with co-solute particles via a second SW attraction of width $\Delta=0.35\sigma_s$ and depth $\varepsilon$. The other colloid($C_2$) interacts via a HS repulsion with the co-solute. This setting provides a realization of $(+,-)$ BC.}
\end{figure}

\begin{figure}[ht]
\centerline{\includegraphics[width=.9\linewidth]{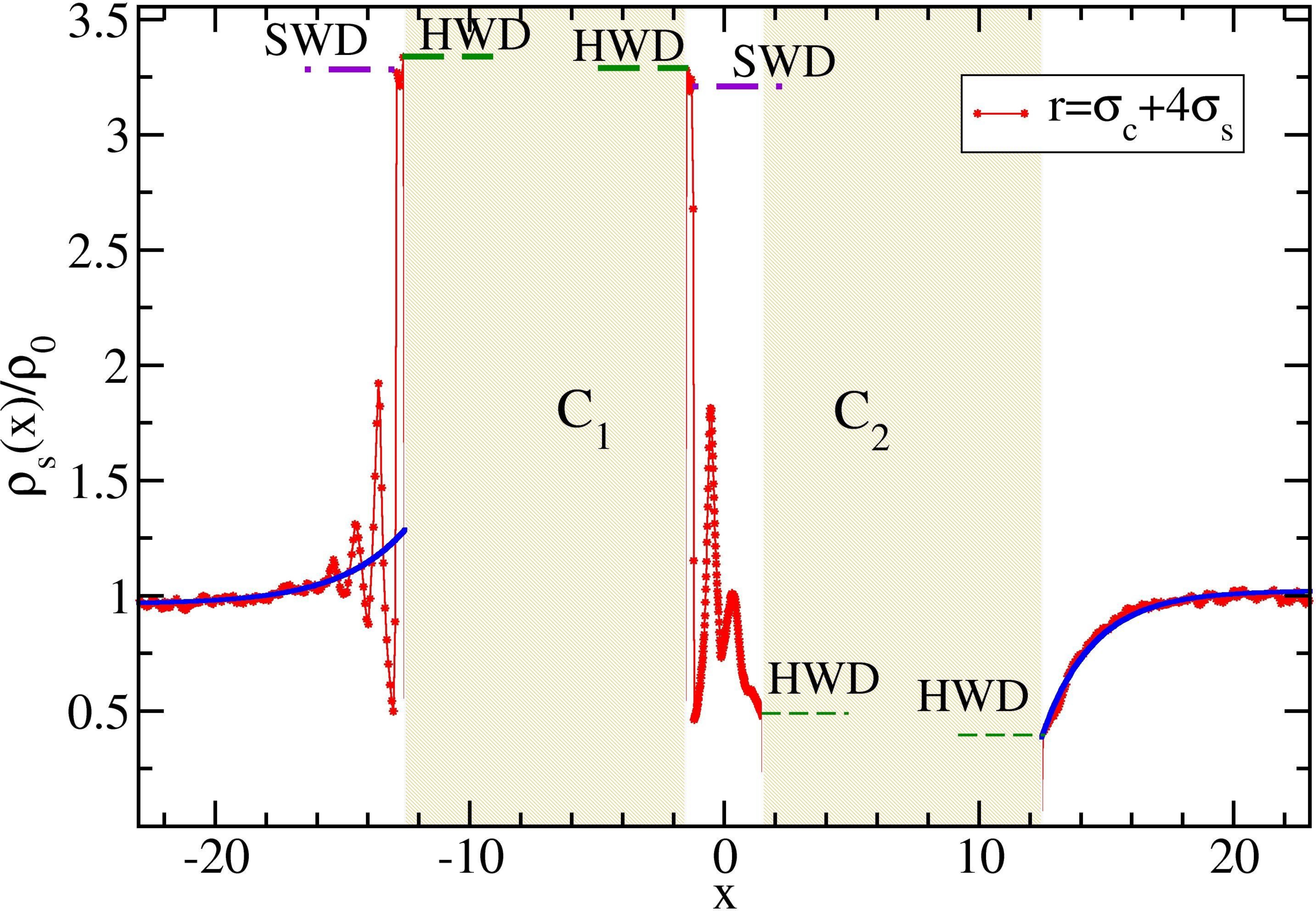}}
\caption{\label{fig7}
Normalized co-solute density profile  with respect to the bulk density $\rho_0$ along the horizontal $x-$axis at $T/T_c=1.0251$ and at the critical density $\phi_c=0.25$.
Co-solute particles interact with a short-range SW attraction of width $\delta=0.1\sigma_s$ and depth $\varepsilon_s$. One of the two colloids ($C_1$) interacts with co-solute particles via a SW attraction of width $\Delta=0.35\sigma_s$ and depth $\varepsilon=1\varepsilon_s$. The other colloid($C_2$) interacts via a HS repulsion with the co-solute. The profile is associated to the effective potential in Fig.~\ref{fig6} (squares).
The different colloid-cosolute interaction results in a different density profile around the two colloids.  Around $C_2$ (colloid on the right), the solvent density is lower than the average, while the opposite behavior is observed on $C_1$ (colloid on the left).  The repulsive force $C_2$ colloid originates  from the slight mismatch of the  contact density (larger inside than outside), while the repulsive force on $C_2$  originates from larger density at the square-well location on the outside, compared to the inside.
The exponential fits (solid lines) have been done by setting the exponential decay to the value of the bulk critical correlation length $\xi=2.5\sigma_s$.}
\end{figure}

Close to the critical point,  the theory predicts that, on increasing $\varepsilon$, $V_{eff}$
changes from an attractive to a repulsive exponential decay, on changing the boundary conditions from
$(-,-)$ to $(-,+)$.  The effective potentials for different $\varepsilon$ close to the critical point are plotted in Fig.~\ref{fig6}.
While for $r<1.3 \sigma_c$ the echo of the layering effects is still visible, for larger $r$ values the decay of all curves is compatible with the same exponential function decay, again supporting the identification of the interaction potential in this spatial region as arising from the universal behavior imposed by the critical fluctuations. 
As in the case of $(-,-)$ BC, for $(+,-)$ BC the density profile provides useful informations on the mechanism that gives rise to the repulsion. Fig.~\ref{fig7} shows the co-solute particles density profile along the $x-$axis. The different colloid-cosolute interaction result in a different density profile around the two colloids.  Around $C_2$ ($-$ BC), the solvent density is lower than the average, while the opposite behavior is observed for $C_1$ ($+$ BC).  The repulsive force on $C_2$ originates from the slight mismatch of the  contact density (larger inside than outside), while the repulsive force on $C_1$ originates from larger density at the SWD on the outside, compared to the inside. 
Contrary to what observed in the $(-,-)$ to $(+,+)$ BC case, here tuning the attraction strength allows us to modify the sign of the effective force~\cite{hertleinnat451}. In fact we observe that for small $\varepsilon$ the effect of the colloid-cosolute attraction is not sufficiently strong to substantially  change the shape of the effective potential,  which remains completely attractive. For higher values of $\varepsilon$ the potential turns into a completely repulsive one and no sign of oscillations driven by co-solute structures is visible. 
It is interesting to note that under these conditions it is in principle possible to tune finely $\varepsilon$ in order to obtain a flat co-solute density profile (apart from the layering at contact). This corresponds to impose Dirichlet BC~\cite{MaciolekPRE79} and from a conceptual point of view to identify a sort of $\Theta$ condition, in analogy with polymer solutions~\cite{Likos}, where the effective interaction potential is close to zero.
\subsection{From the $(+,+)$ to the $(+,-)$ BC} 
\begin{figure}[ht]
\centerline{\includegraphics[width=.9\linewidth]{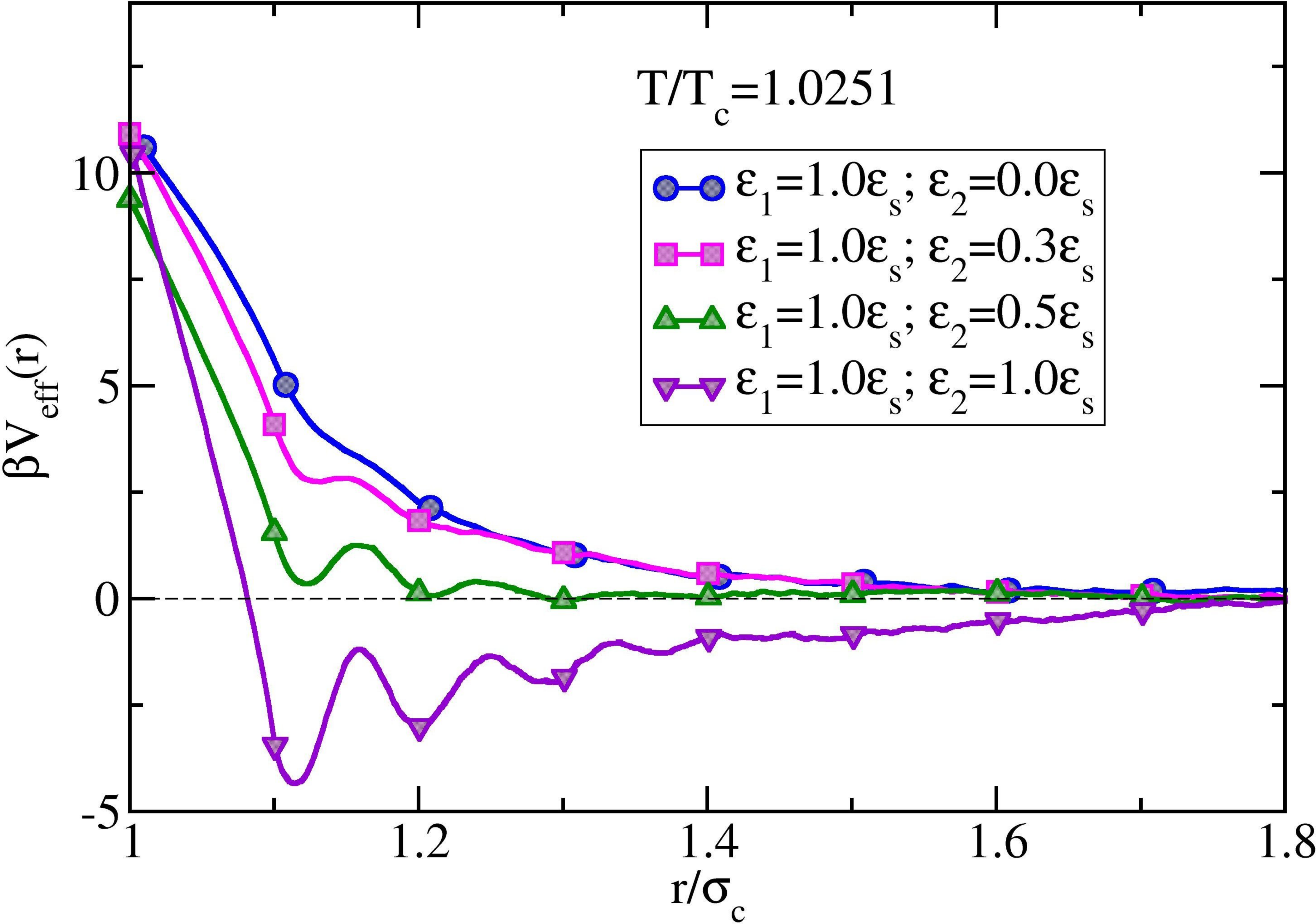}}
\caption{\label{fig8}
Evolution of the effective potential for two colloids close to the critical point of the co-solute ($T/T_c=1.0251$ and $\phi=\phi_c=0.25$). One of the two colloids ($C_1$) interacts with co-solute particles via a SW attraction of width $\Delta=0.35\sigma_s$ and depth $\varepsilon_1=1.0\varepsilon_s$. The other colloid ($C_2$) interacts with the co-solute through a SW attraction as well with the same width $\Delta$ but depth $\varepsilon_2$. Starting from $\varepsilon_2=\varepsilon_1$, i.e. $(+,+)$ BC, the interaction strength $\varepsilon_2$ is decreased down to the limit case of $\varepsilon_2=0$ corresponding to $(+,-)$ BC. Contrary to the case illustrated in Fig.~\ref{fig6}, here only the long distance tail can be tuned from attractive to repulsive, while the short range part of the potential remains repulsive.}
\end{figure}

\begin{figure}[ht]
\centerline{\includegraphics[width=.9\linewidth]{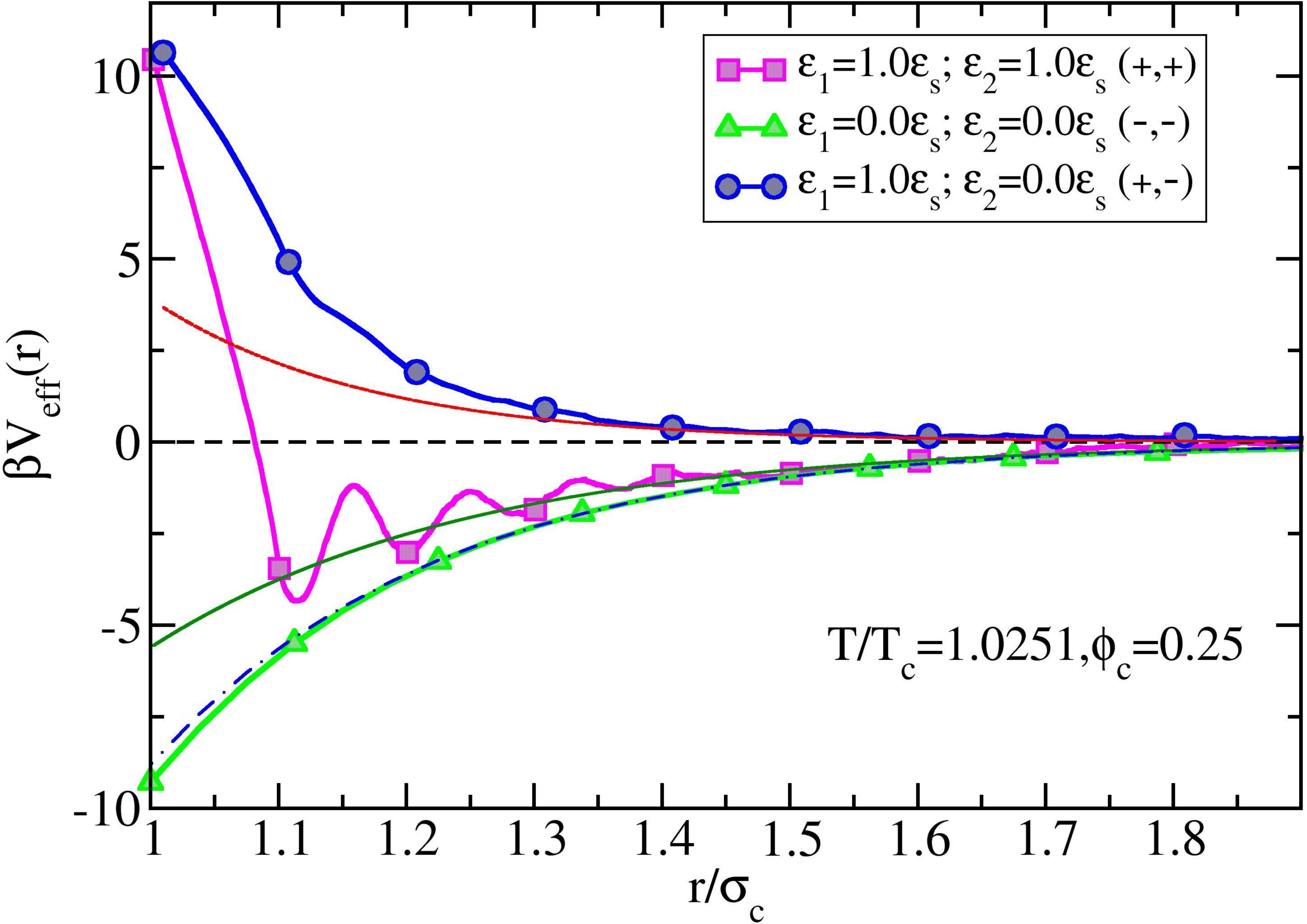}}
\caption{\label{fig9} Effective potentials for different BC, evaluated at the co-solute critical packing fraction and at $T/T_c=1.0251$. In the critical region, independently on the BC, the long range part of the effective potential decays exponentially. According to eq. (\ref{eq:Phi_z}) it is possible to extract the characteristic length  that controls the decay of the exponential. To extract this value the three curves have been interpolated in the same range with eq. (\ref{eq:Phi_z}) starting from $r=1.3\sigma_c$. For the three cases we find $\xi_{(-,-)}=2.24\sigma_s$, $\xi_{(+,+)}=2.49\sigma_s$, and $\xi_{(+,-)}=2.20\sigma_s$ which is consistent the bulk correlation length $\xi=2.5\sigma_s$ of the co-solute for this $T/T_c$~\cite{GnanSM8}. The non universal amplitudes found from the exponential fits are $A_{(-,-)}=-0.63$ $A_{(+,+)}=-0.45$ $A_{(+,-)}=0.48$.} 
\end{figure}

According to theoretical predictions~\cite{gambassipre80} the effective potential is expected to change in the transition from $(+,+)$ to $(+,-)$ exactly as in the case from $(-,-)$ to $(+,-)$.  These predictions refer to the scaling region of the potential.
In the present numerical study, we also access the short distances, where the effective potential probes the
non-universal aspects of the solvent-colloid interaction.  To highlight the difference at short distances between the two cases, Fig.~\ref{fig8} shows the evolution of the effective potential when the attraction between only one of the two colloids and the co-solute is progressively reduced.  Differently from the results of Fig.~\ref{fig6}, in this case only the long distance behavior changes sign, while the short distance part of the potential remains always repulsive.
Finally in  Fig.~\ref{fig9} we plot cases differing in their BCs, but all at the same temperature and co-solute critical packing fraction.  In all cases, the long-distance behavior of $V_{eff}$ can be described, as expected theoretically, by an exponential decay, with a correlation length  consistent with the bulk critical correlation length $\xi$.
\section{Conclusions}
In this article we have discussed how the effective potential, resulting from confining critical co-solute particles between the surfaces of two large colloids, depends on  the colloid-cosolute interaction.  
In a previous study~\cite{GnanSM8} we have already investigated the case of $(-,-)$ BC, i.e. when only colloid-cosolute hard-core repulsion is present. In such case we have shown that the resulting potential for $T/T_c\rightarrow 1$ at the co-solute critical packing fraction is monotonic, attractive and long range. The co-solute density profile along the $x-$axis of the simulation box shows that for $(-,-)$ BC the density close to colloids is lower than the bulk density and it relaxes exponentially to the bulk value far from the two colloids.
Upon switching on the colloid-cosolute attraction $\varepsilon$ , depletion effects are progressively weakened and the co-solute density close to the colloid varies from values smaller than the bulk to values larger than the bulk. Correspondingly,  the contact value of the potential varies from negative to positive while the long distance part of $V_{eff}(r)$, dominated by critical fluctuations, remains always attractive  as theoretically predicted in the $(+,+)$ BC case.   The increase in $\varepsilon$  has instead a  profound effect on the non-universal short-distance part of the potential,  which 
progressively develops large oscillations. Such oscillations are related to specific geometries 
at characteristic lengths associated to integer number of co-solutes between the two colloids. At these
specific distances, the local energy is minimized.   For intermediate values of $\varepsilon$  the colloid-cosolute
attraction  compensates the depletion interaction, effectively reducing the short-distances interaction potential.
We have also addressed the case of two colloids interacting differently with the co-solutes,
a realization of the so-called  $(+,-)$ BC case.  In this case, the local density close to the
two colloids is respectively lower and higher than the bulk value and the critical Casimir forces
are expected to be repulsive.  We have shown that indeed, when the asymmetry in the interaction potential
is sufficiently intense to drive different BC,  the potential becomes repulsive at all length scales. 
We have shown that, independently on the BCs,  the critical long range part of the effective potential
is always described by an exponential whose decay is controlled by the critical correlation length $\xi$ corresponding to the thermal correlation length of the bulk co-solute close to the critical point, in full agreement with theoretical predictions~\cite{gambassipre80}. 
We have investigated the behavior of the density profile in all examined BC cases.  Building on the
fact that all interactions are modeled as excluded volume or as square-well attraction, the net pressure over the
colloids can be estimated simply by the density of co-solute at contact and at the square-well distance (in the case of attraction between the colloid and the cosolute). The density profiles confirm that in the $(-,-)$ case, the
contact density inside is smaller than outside, determining the net attraction. In the $(+,+)$ case a different mechanism for attraction is observed:  the leading contribution arises from the mismatch of the density at the well distance, larger  inside than outside, determining a net attraction. Finally, in the $(+,-)$ case, repulsion is 
driven by two different mechanisms for the two colloids. The hard-sphere colloid is pushed out by the
larger contact density inside. The attractive colloid is dragged out by a larger density at the square well distance.

The possibility of varying the potential from repulsive to attractive and to finely control its shape and intensity by tuning the BC, provides a guidance for controlling equilibrium properties of colloids dispersed in precritical suspensions. This is important for future applications, for instance in the case in which the confining surfaces are chemically patterned colloidal particles, such as patchy~\cite{SciortinoCurr2011} or Janus particles~\cite{janus-softmatter,GambassiSM7}. In this case, the introduction of a geometrical constraint in the colloid-cosolute interaction would give rise to a torque that can be used to control the orientation of colloids, as shown for chemically patterned substrates~\cite{soykaprl101}, and that could give rise to new unexplored phases. More in general, our results provides useful informations for designing interaction potentials which can be exploited to control the stability of colloidal systems.

\section{Acknowledgments:} We acknowledge support from MIUR-PRIN Project 2008 CX7WYL and ERC-226207-PATCHYCOLLOIDS.  
%\bibliography{biblio_patchy}
\bibliographystyle{empty}

\end{document}